\documentclass[letterpaper,10pt,conference]{ieeeconf}

\IEEEoverridecommandlockouts
\usepackage{amsmath,amssymb,mathrsfs}
\usepackage{graphicx}
\usepackage{subfig}
\usepackage{color}
\usepackage{bm}
\usepackage{amsfonts}
\usepackage{latexsym}
\usepackage{url}
\usepackage{ifthen}
\usepackage{comment}
\usepackage{multirow}

\newtheorem{theorem}{Theorem}

\newcommand{\ket}[1]{\left|#1\right\rangle}
\newcommand{\bra}[1]{\left\langle#1\right|}

\newcommand{\OUT}{\mathrm{OUT}}
\newcommand{\IN}{\mathrm{IN}}

\newcommand{\norm}[1]{|| #1 ||}

\def\Tr{\operatorname{Tr}}
\renewcommand{\H}{\mathbf{H}}
\newcommand{\X}{\mathbf{X}}
\newcommand{\Y}{\mathbf{Y}}
\newcommand{\Z}{\mathbf{Z}}
\newcommand{\Id}{\mathbf{I}}
\renewcommand{\L}{\mathbf{L}}
\newcommand{\D}{\mathbf{D}}
\renewcommand{\S}{\mathbf{S}}
\newcommand{\HD}{\mathbf{H_D}}
\newcommand{\VD}{\mathbf{V_D}}

\newcommand{\V}{\mathbf{V}}
\newcommand{\K}{\mathcal{K}}

\renewcommand{\vec}{\bm}

\usepackage{tikz}
\newcommand\copyrighttext{\footnotesize \textcopyright 2018 IEEE. Personal use of this material is permitted.
  Permission from IEEE must be obtained for all other uses, in any current or future
  media, including reprinting/republishing this material for advertising or promotional
  purposes, creating new collective works, for resale or redistribution to servers or
  lists, or reuse of any copyrighted component of this work in other works.
}
\newcommand\copyrightnotice{%
\begin{tikzpicture}[remember picture,overlay]
  \node[anchor=north,yshift=-10pt] at (current page.north)
       {\fbox{\parbox{\dimexpr\textwidth-\fboxsep-\fboxrule\relax}{\copyrighttext}}};
\end{tikzpicture}}

\title{Robustness of energy landscape control for spin networks under decoherence}
\author{S.\ Schirmer$^{1}$, E.\ Jonckheere$^{2*}$, S.\ O'Neil$^3$, and F.\,C.\ Langbein$^{4}$
\thanks{$^{1}$ Swansea University, UK, \texttt{lw1660@gmail.com}}
\thanks{$^{2*}$ U.~Southern California, \texttt{jonckhee@usc.edu}, corresponding author}
\thanks{$^{3}$ US Mil.~Acad., \texttt{Sean.O'Neil@usma.edu}}
\thanks{$^{4}$ Cardiff University, UK, \texttt{frank@langbein.org}}}

\begin{document}
\enlargethispage{0.5in}
\maketitle
\copyrightnotice

\begin{abstract}
Quantum spin networks form a generic system to describe a range of quantum devices for quantum information processing and sensing applications. Understanding how to control them is essential to achieve devices with practical functionalities. Energy landscape shaping is a novel control paradigm to achieve selective transfer of excitations in a spin network with surprisingly strong robustness towards uncertainties in the Hamiltonians. Here we study the effect of decoherence, specifically generic pure dephasing, on the robustness of these controllers. Results indicate that while the effectiveness of the controllers is reduced by decoherence, certain controllers remain sufficiently effective, indicating potential to find highly effective controllers without exact knowledge of the decoherence processes.
\end{abstract}

\section{Introduction}

The promise of applications ranging from quantum computing to metrology has resulted in strong interest in coupled spin systems, or spin networks for short, as potential prototype systems for quantum information processing and sensing applications~\cite{QuantumSpintronicsReview}. As control plays a fundamental role in the translation of physical phenomena into technology, the development and implementation of effective control schemes for quantum systems are essential to harness the technological potential of quantum systems~\cite{Glaser2015}.  Much of the quantum control literature has focused on dynamic control of the system Hamiltonian via time-varying external control fields.  Recently, an alternative paradigm for quantum control based on energy landscape shaping has been proposed and applied to derive feedback control laws for selective transfer of excitations between nodes in a spin network~\cite{Edmond_IEEE_AC}.

Controllers $D(\ket{\IN},\ket{\OUT},T)$ are designed to maximize the fidelity of transfer from the input $\ket{\IN}$ to the output $\ket{\OUT}$ at a specified readout time $T$ or time window $[T-\delta T,T+\delta T]$, using only static fields to shift the energy levels of the system~\cite{time_optimal}. Previous work considered the ideal case of coherent transport of systems subject to unitary evolution. While this assumption can be justified for systems whose dynamics are restricted to a decoherence-free subspace, or which are sufficiently well-isolated from their environment to render decoherence due to unwanted interactions negligible on the timescales of interest, most quantum systems are affected by decoherence.

In this paper we study the effect of pure dephasing on the effectiveness of energy-landscape shaping control in spin networks.  In Sec.~\ref{sec:theory} the theory of quantum spin networks and their evolution under decoherence is introduced, followed by a brief summary of the dynamic regimes and control objectives in Sec.~\ref{sec:control}.  The main results on the sensitivity of the transfer fidelity in the presence of decoherence are presented in Sec.~\ref{sec:robustness}.

\section{Spin Networks subject to Decoherence} \label{sec:theory}

\subsection{Spin network Hamiltonian}

A network of $N$ interacting spin-$\tfrac{1}{2}$ particles with near neighbor couplings and bias fields can be described by a $2^N \times 2^N$ Hamiltonian of the form $\H_{\D} = \H + \D$, where
\begin{subequations}
  \begin{align}
    \H &= \sum_{(m,n) \in \mathcal{E}} J_{mn} (\X_m\X_{n}+\Y_m\Y_{n} +\kappa \Z_m\Z_{n}),\\
    \D &= \sum_{n=1}^N D_n \Z_n.
\end{align}
\end{subequations}
$\X_n$, $\Y_n$, $\Z_n$ are Pauli spin operators acting on spin $n$, i..e, $N$-fold tensor products whose $n$th factor is
\begin{equation*}
  X = \begin{pmatrix} 0 & 1 \\ 1 & 0 \end{pmatrix},
  Y = \begin{pmatrix} 0 & -\jmath \\ \jmath & 0 \end{pmatrix},
  Z = \begin{pmatrix} 1 & 0 \\ 0 & -1 \end{pmatrix},
\end{equation*}
respectively, all other factors being the $2\times 2$ identity matrix $I$.  $\kappa$ is used to distinguish different interaction types, e.g., XX coupling ($\kappa=0$) and Heisenberg coupling ($\kappa=1$).  $J_{mn}$ denotes the strength of the coupling between the $m$th and $n$th node and  $D_n$ the static bias field at spin $n$.   $\mathcal{E}$ denotes the set of edges in the corresponding graph associated with the network.

\subsection{Evolution under decoherence}

The state of the system at time $t$ can be described by a $2^N \times 2^N$ density operator $\varrho(t)$. If the system is weakly coupled to an environment, then the evolution can generally be described by a Lindblad equation
\begin{equation}\label{e:QME1}
  \dot{\varrho}(t) = -\jmath [\H_\D,\varrho(t)] + \L_\D(\varrho(t)),
\end{equation}
where $\HD$ is the Hamiltonian defined above and $\L_\D$ is a Lindblad super-operator
\begin{equation}\label{e:QME2a}
  \L_\D(\varrho) =  \VD \varrho \VD^\dag  - \tfrac{1}{2} (\VD^\dag \VD \varrho  + \varrho \VD^\dag \VD).
\end{equation}
For $\VD=0$ we recover the usual Hamiltonian dynamics considered in previous work \cite{Edmond_IEEE_AC}.  In this paper we are mostly interested in systems subject to decoherence, which can be modeled as dephasing in the Hamiltonian basis and described by Lindblad operators $\L_\D$ of dephasing type, given by Hermitian dephasing operators $\VD$ that commute with the system Hamiltonian, $[\HD,\VD]=0$.  $\VD=\VD^\dag$ further implies that Lindblad superoperator can be simplified to
\begin{equation}\label{e:QME2}
  \L_\D(\varrho) = - \tfrac{1}{2} [\VD, [\VD, \varrho]].
\end{equation}
The subscript $\D$ here indicates dependence on the control as strictly speaking decoherence in the weak coupling limit depends on the total Hamiltonian and hence the control~\cite{domenico_CDC,singular_vs_weak_coupling}. Although simple, this model is closer to  the master equation in  the weak coupling limit developed in~\cite{singular_vs_weak_coupling} as it appears at a first glance.

As $\H_\D$ and $\VD$ commute, they are simultaneously diagonalizable and there exists a set of projectors $\{\Pi_k(\HD)\}_{k}$ onto the (orthogonal) simultaneous eigenspaces of $\HD$ and $\VD$ such that $\sum_{k}\Pi_k(\HD)=\Id_{\mathbb{C}^{2^N}}$ is a resolution of the identity on the full Hilbert space $\mathbb{C}^{2^N}$ and
\begin{align*}
  \H_\D = \sum_{k} \lambda_k(\HD) \Pi_k(\HD), \quad
  \VD   = \sum_{k} c_k \Pi_k(\HD) ,
\end{align*}
where $\lambda_k(\HD)$ and $c_k$ are the real eigenvalues of $\HD$ and $\VD$, respectively.  Pre-/post-multiplying the master equation~\eqref{e:QME1} with Lindblad term \eqref{e:QME2} by  $\Pi_k(\HD)$ and $\Pi_\ell(\HD)$, respectively, yields
\begin{equation} \label{e:rho_k_ell}
  \Pi_k(\HD)\dot{\varrho}(t) \Pi_{\ell}(\HD)= (-\jmath \omega_{k\ell}+\gamma_{k\ell}) \Pi_k(\HD) \varrho(t) \Pi_\ell(\HD),
\end{equation}
with $\omega_{k\ell}= \lambda_k-\lambda_\ell$ and $\gamma_{k\ell}=-\tfrac{1}{2}(c_k-c_\ell)^2\le 0$ and solution
\begin{equation*}
  \Pi_k(\HD)\varrho(t)\Pi_\ell(\HD) = e^{-t(\jmath \omega_{k\ell}-\gamma_{k\ell})} \Pi_k(\HD) \varrho_0 \Pi_\ell(\HD).
\end{equation*}
The above clearly shows the decoherence $\gamma_{k\ell}$ acting on the subspace $\Pi_k(\HD)\varrho \Pi_\ell(\HD)$. Since $\sum_k \Pi_k(\HD) = \Id$, the full solution is found as  $\varrho(t)=\sum_{k,\ell}\Pi_k(\HD)\varrho(t)\Pi_\ell(\HD)$, which gives explicitly
\begin{equation}
 \varrho(t) = \sum_{k,\ell} e^{-t(\jmath \omega_{k\ell}-\gamma_{k\ell})} \Pi_k(\HD) \varrho_0 \Pi_\ell(\HD).
\label{e:varrho_of_t_solution}
\end{equation}

\subsection{Subspace Dynamics}

The total Hamiltonian $\H_\D$ commutes with the operator
\begin{equation}
  \S= \frac{1}{2} \sum_{n=1}^N (\Id +\Z_n),
\end{equation}
which counts the number of excited spins in the network.   Therefore, $\H_{\D}$ and $\S$ have the same eigenspaces, and for decoherence acting in the Hamiltonian basis, the dynamics of each excitation subspace remains effectively decoupled.  We can therefore restrict our attention to individual subspaces, as considered in previous work on coherent transport.  In particular we can retain the eigenspace of $\S$ corresponding to the eigenvalue $1$, often referred to it as the \emph{single excitation subspace.}  $[\HD,\S]=0$ implies that the single excitation subspace is composed of eigenspaces of $\HD$.  Let $\mathcal{K}$ be the set of indexes of eigenspaces of $\HD$ that span the single excitation subspace, and define the single-excitation subspace operators
\begin{subequations}\label{e:total_to_reduced}
\begin{align}
  V_D &= \sum_{k \in \K}c_k \Pi_k(\HD), \\
  H_D &=\sum_{k \in \K} \lambda_k \Pi_k(\HD),\\
  \rho &=\sum_{k,\ell \in \K}\Pi_k(\HD) \varrho \Pi_\ell(\HD).
\end{align}
\end{subequations}
Then the reduced Lindblad-Liouville equation is
\begin{equation}
  \dot{\rho}=-\jmath [H_D,\rho]+ V_D \rho V_D  - \tfrac{1}{2} (V_D^2 \rho + \rho V_D^2)
\label{e:QME1r}
\end{equation}
and Eq.~\eqref{e:varrho_of_t_solution} shows that the solution is
\begin{equation}
\label{e:rho_of_t_solution}
  \rho(t) = \sum_{k,\ell \in \K} e^{-t(\jmath \omega_{k\ell}-\gamma_{k\ell})} \Pi_k(H_D) \rho_0 \Pi_\ell(H_D).
\end{equation}
Eq.~\eqref{e:rho_of_t_solution} provides a computationally efficient way to simulate the dynamics of the spin network subject to dephasing for a given system and controller.

\textit{Remark.}  In the case of collective dephasing, $\V=\S$.  Thus, $c_k$ are the eigenvalues of $\S$, with $c_k=1$ for the single excitation subspace, and the subspace is decoherence free~\cite[Sec. III.B]{collective_dephasing}.

\subsection{Special network topologies}

For special spin networks with a simple topology such as a ring or a chain this procedure yields the single excitation subspace Hamiltonian for the controlled system
\begin{equation}
\label{e:Hamiltonian}
H_D = \begin{pmatrix}
D_1      & J_{12} & 0       & \ldots  & 0    & J_{1,N} \\
J_{12}   & D_2    & J_{23}  &                & 0 & 0\\
0        & J_{23} & D_3     &                & 0 & 0\\
 \vdots  &        &  \ddots &  \ddots        & \ddots& \vdots\\
 0       & 0      &  0      &      & D_{N-1} & J_{N-1,N}\\
J_{1,N}  & 0      &   0     &  \ldots        & J_{N-1,N} & D_{N}
\end{pmatrix}
\end{equation}
with $J_{1N}=0$ for a chain.   Perturbations of the $J_{mn}$ values do not change the structure of $\H_\D$ and hence do not affect the commutativity relation between $\H_\D$ and $\S$.  Similarly, errors in the field focusing only affect the $D_n$ values and hence do not affect the structure of $\H_\D$. Therefore, invariance of the single excitation subspace is maintained under both near-neighbor coupling strength and bias field control perturbations.

\section{Dynamic Regimes and Control Objectives} \label{sec:control}

\subsection{Objectives}

\begin{figure}\center
  \scalebox{0.6}{\includegraphics[viewport=200 130 450 380]{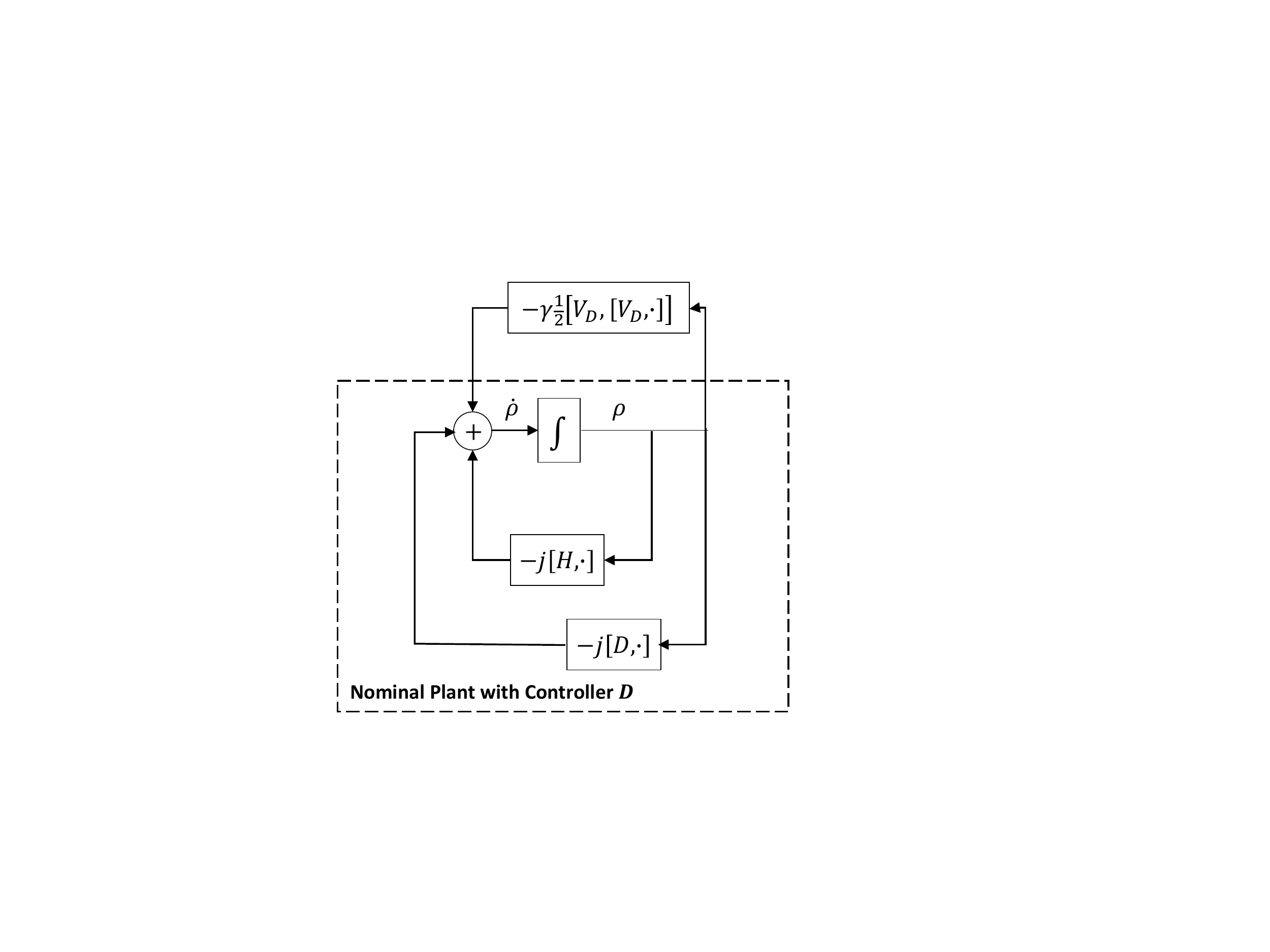}}
  \caption{Nominal plant with controller}\label{fig:sys}
\end{figure}

The nominal plant and controller setup is shown in Fig.~\ref{fig:sys}. As in previous work, our control objective is to find a controller that steers the dynamics to maximize the transfer fidelity of a local excitation at one node of the spin network, $\ket{\IN}$, to another node, the output node $\ket{\OUT}$.  Depending on the application, we consider either the instantaneous transfer fidelity at a certain time $T$,
\begin{equation}
    p(\IN\to\OUT, T) =  \bra{\OUT} \rho(T) \ket{\OUT},
\end{equation}
the time-average over a certain readout time window,
\begin{equation}
  \bar{p}(\IN\to\OUT, T,\delta T) = \frac{1}{2\delta T} \int_{T-\delta T}^{T+\delta T}\!\!\! \bra{\OUT} \rho(t) \ket{\OUT}\, dt,
\end{equation}
where $\rho(t)$ is the solution of Eq.~\eqref{e:QME1r} with $\rho(0)=\ket{\IN}\bra{\IN}$, or long-term average transfer
\begin{equation}
  \overline{p}(\IN\to\OUT, 0, \infty) = \lim_{T \to \infty} \frac{1}{T}\int_0^{T} \bra{\OUT}\rho(t)\ket{\OUT} \, dt.
\end{equation}
The latter is particularly useful for long-term localization if $\IN =\OUT$.

The transfer error is given by $1-p(\IN\to\OUT,T)$, $1-\bar{p}(\IN\to\OUT, T,\delta T)$ or $\overline{p}(\IN\to\OUT, 0, \infty)$, respectively.

Finding globally optimal controllers is computationally expensive due to the complex optimization landscape. The controllers used here were calculated with a restart L-BFGS algorithm, requiring many restarts to find high-fidelity controllers~\cite{time_optimal}.

\subsection{Dynamic regimes}

The most common and non-classical regime is coherent unitary dynamics. In this case the eigenstates of the Hamiltonian are steady states but the dynamics are oscillatory. Although it is possible to design dynamic feedback control laws that render certain eigenstates of the system attractive~\cite{Wang2010}, for a fixed control law, unitary evolution implies that no trajectories converge to a steady state and therefore there are no asymptotically stable states. The best we can do is to seek a controller that maps a desired input state to a desired output state or implements a desired unitary operation. This is the framework adopted in earlier work~\cite{Edmond_IEEE_AC,time_optimal}.

At the other extreme, in the strong backaction or strongly dissipative regime, there are asymptotically stable, globally attractive steady states~\cite{Schirmer2010}. Although the behavior in terms of robustness and stability is more classical, it is interesting to note that we can still stabilize highly non-classical states, e.g., entangled states under the right conditions~\cite{Motzoi2016}.

Our focus here is on the intermediate regime of coherent dynamics with dephasing in the eigenbasis of the system; we have neither asymptotically stable states nor fully coherent dynamics but oscillatory behavior damped by dephasing.  This leads to the emergence of steady states, which are typically classical mixed states, that form a manifold of steady states and are not asymptotically stable~\cite{Schirmer2010}.  If the Hamiltonian and decoherence operators are fixed, the steady state we converge to depends on the initial state, while the rate of convergence to the steady state depends on the degree of phase damping.

If decoherence is sufficiently weak such that we remain far away from a steady state then the dynamics are dominated by non-equilibrium, mostly coherent transport.  This case can be treated as a perturbation of  the unitary evolution case.  If the phase damping is strong relative to the transfer time then the system will approach a steady state and the probability of the transport will be determined by the overlap of the output state with the steady states of the system.

\subsection{Long-term time-averages and asymptotic steady states}

Looking at Eq.~\eqref{e:rho_of_t_solution} under coherent dynamics ($\gamma_{k\ell}=0$) and noting that $\omega_{kk}=0$, it is obvious that the solution $\rho(t)$ oscillates around $\sum_{k\in \mathcal{K}} \Pi_k(H_D)\rho_0\Pi_k(H_D)$ and therefore has a \emph{long-term time-average}
\begin{equation}\label{e:long_term}
  \lim_{T \to \infty} \frac{1}{T} \int_0^T \langle\mathrm{OUT}|\rho(t)|\mathrm{OUT}\rangle dt
  =  \langle \mathrm{OUT}|\rho_\infty \mathrm{OUT}|\rangle,
\end{equation}
where
\begin{equation}
  \rho_\infty = \sum_{k\in\mathcal{K}} \Pi_k(H_D) \rho_0 \Pi_k(H_D).
\end{equation}
(A rigorous proof can be constructed from the \emph{nonclassical} Laplace final value theorem~\cite[Th. 2]{formal_extended_final_value_Laplace}).

On the other hand, in the dephasing case, if the dephasing rates $\gamma_{k\ell} \ne 0$ for $k \ne \ell$, all terms with $k\ne \ell$ on the right-hand side of Eq.~\eqref{e:rho_of_t_solution} vanish as $t \to \infty$, while the $k=\ell$ terms survive.  Therefore, any initial state $\rho_0$ converges to a \emph{steady state}
\begin{equation}\label{e:steady_state}
  \begin{split}
    \lim_{t \to \infty} \bra{\OUT}\rho(t)\ket{\OUT}
    &= \bra{\OUT} \rho_\infty \ket{\OUT} \\
    &= \Tr[\rho_{\OUT} \rho_\infty].
  \end{split}
  \end{equation}

Comparing Eqs.~\eqref{e:long_term} and~\eqref{e:steady_state}, it follows that the steady states for the dephasing system can be related, via $\rho_\infty$, to the long-term time-averaged states for the fully coherent case.

Maximizing the asymptotic transfer fidelity~\eqref{e:steady_state} is equivalent to maximizing the overlap $\Tr[\rho_{\OUT} \rho_\infty]$ of the target density operator $\rho_{\OUT}$ with the steady state of the system.  The asymptotic transfer fidelity, and therefore the long-term average transfer fidelity, depend on the control through the control dependence on the projectors $\Pi_k(H_D)$.   In the special case where input and target states are pure states, inserting $\rho_0 = \ket{\IN}\bra{\IN}$ and $\rho_{\OUT}=\ket{\OUT}\bra{\OUT}$ yields
\begin{equation}
  \Tr[\rho_{\OUT} \rho_\infty]  = \sum_{k\in\mathcal{K}} |\bra{\OUT} \Pi_k(H_D) \ket{\IN}|^2.
\end{equation}

Maximizing the long-term average or asymptotic transfer fidelity is therefore equivalent to maximizing the sum of the squares of the mutual overlaps of the initial and target states with the eigenspaces of the Hamiltonian, or the $L_2$ norm of $\vec{y}$ with  $y_k = \bra{\OUT} \Pi_k(H+D)  \ket{\IN}$.  Notice the similarily to Eq.~(18) in \cite{Edmond_IEEE_AC} and the necessary condition for superoptimality (the controller achieves perfect state transfer at some time $T$ in the coherent case), which requires that the supremum over all controllers of
\begin{equation}
 \sum_{k \in \mathcal{K}} |\bra{\OUT} \Pi_k(H+D)  \ket{\IN}|,
\end{equation}
i.e., the $L_1$ norm of $\vec{y}$, reaches its upper bound of $1$.  In this case, however, achieving the upper bound of the $L_1$ norm of $\vec{y}$ is only a necessary condition and there is a second ``phase matching'' condition required for optimality.  The loss of the phase matching condition in the presence of dephasing or long-term averaging makes sense as phase information is lost as a result of dephasing or taking long term averages.

\section{Sensitivity of transfer fidelity} \label{sec:robustness}

\subsection{Asymptotic transfer probability}

When the transfer time is long compared to the time required for the system to reach a steady state, it is useful to consider the sensitivity of the asymptotic probability of transfer (squared fidelity) $p_\infty =\bra{\OUT}\rho_{\infty }\ket{\OUT}$ and compute the log-sensitivity in the same manner as~\cite{statistical_control}.   Using the perturbed, controlled Hamiltonian ${\widetilde{H}}_D=H+D+\delta S_{H_D}$ where $S_{H_D}$ indicates the (certain) structure of the perturbation and $\delta$ its (uncertain) strength, we have
\begin{equation} \label{eq:pinf}
   p_\infty = \sum_{k \in \mathcal{K}}  \bra{\OUT} \Pi_k(\widetilde{H}_D) \rho_0 \Pi_k(\widetilde{H}_D) \ket{\OUT}
\end{equation}
with $\rho_0 = \ket{\IN}\bra{\IN}$. $p_\infty$ depends on $\delta$ via $\Pi_k(\widetilde{H}_D)$ while $\rho_0$ and $\OUT$ are fixed, so applying the product rule gives
\begin{equation}\label{eq:der}
  \begin{split}
  \frac{\partial p_\infty}{\partial \delta}
       =&  2 \Re \sum_{k \in \mathcal{K}}  \bra{\OUT} \frac{\partial \Pi_k(\widetilde{H}_D)}{\partial \delta} \rho_0 \Pi_k(\widetilde{H}_D) \ket{\OUT}.
  \end{split}
\end{equation}
$\frac{\partial p_\infty}{\partial \delta}$ provides a measure of the sensitivity of the asymptotic fidelity to a parameter variation of size $\delta$ structured as  $S_{H_D}$.

To calculate $\frac{\partial \Pi_k(\widetilde{H}_D)}{\partial \delta}$  we assume that $\Pi_k$ is the projector onto a 1D eigenspace, $\Pi_k(\widetilde{H}_D) = \ket{v_k}\bra{v_k}$, so that
\begin{equation} \label{eq:der2}
  \frac{\partial \Pi_k (\widetilde{H}_D)}{\partial \delta} = \ket{ \frac{\partial v_k}{\partial \delta}} \bra{v_k}
  + \ket{v_k} \bra{\frac{\partial v_k}{\partial \delta}},
\end{equation}
where $v_k$ are the eigenvectors of $\widetilde{H}_D$.  We then calculate the derivatives of the eigenvectors in accordance with~\cite{new_eig_derivatives, van_der_Aa}.

Interpreting $\epsilon_\infty=1-p_\infty$ as an error term, and inserting \eqref{eq:der2} into \eqref{eq:der}, yields the following expression for the logarithmic sensitivity of the error
\begin{equation}
   \left| \frac{1}{\epsilon_{\infty }} \frac{\partial \epsilon_{\infty }}{\partial \delta} \right|\\
  = \frac{2 \Re \sum_k \langle \OUT | \tfrac{\partial}{\partial \delta}\Pi_k(\widetilde{H}_D) \rho_0 \Pi_k (\widetilde{H}_D) | \OUT \rangle }{1-\langle \OUT | \rho_{\infty } | \OUT \rangle }.
\end{equation}

This logarithmic sensitivity of the error has been used to assess whether the $D$-controller is ``classical''  or ``anti-classical''  in the sense of conflict or no conflict, resp., between (tracking) error and log-sensitivity to model uncertainties. In~\cite{statistical_control} and~\cite{ssv_mu},  it was shown that for coherent dynamics the $D$-controller is ``anti-classical." Here, recovery of classicality under decoherence is found (see Fig.~\ref{fig:N5_sensitivity1}(b)) based on a \emph{\it large} random sampling of the decoherence processes.

\subsection{Non-asymptotic Regime}

To investigate the sensitivity of the transfer fidelity to weak decoherence in the non-asymptotic regime, we perform simulations for various test cases consisting of rings with XX-coupling, restricted to the single excitation subspace.

For each system a set of time-invariant static bias fields $\vec{D}$ was computed by numerically maximizing the probability of transport for a given transfer time $T$ or time window as described in previous work \cite{Edmond_IEEE_AC} for the ideal Hamiltonians without decoherence using a restart L-BFGS algorithm~\cite{time_optimal}.  Excitation transfers from the initial state $\ket{\IN} = \ket{1}$ to a final state $\ket{\OUT} = \ket{n}$ for $n=1,2,\ldots, \lceil \frac{N}{2}\rceil$ were considered, where $\ket{n}$ denotes an excitation localized at a spin $n$.  For each transfer problem, $1000$ to $2000$ independent controllers were calculated and sorted according to the transfer fidelity achieved in each case, in the absence of decoherence.

\begin{figure} \center
 \subfloat[High fidelity controller]{\includegraphics[width=0.9\columnwidth]{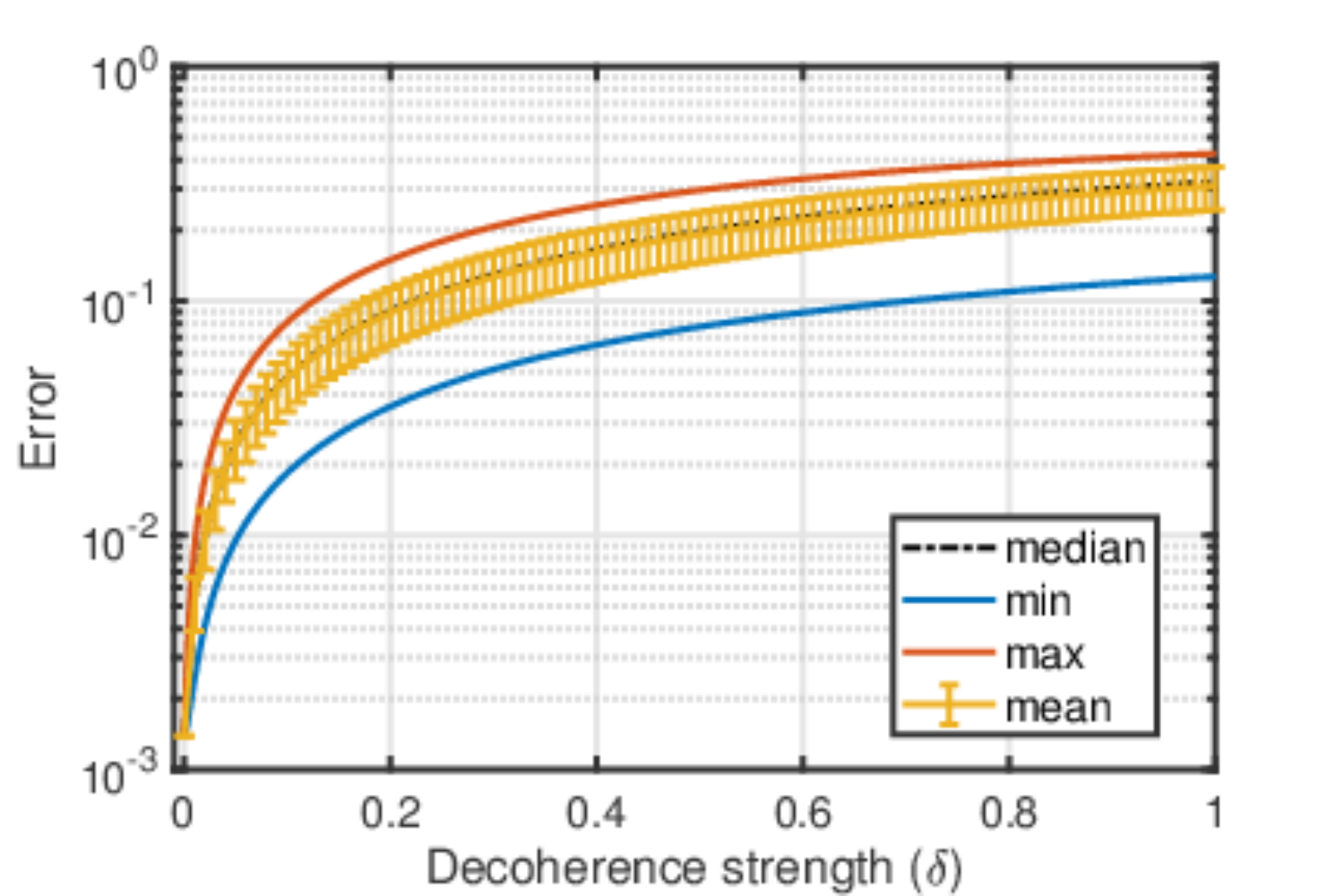}} \\
 \subfloat[Low fidelity controller]{\includegraphics[width=0.9\columnwidth]{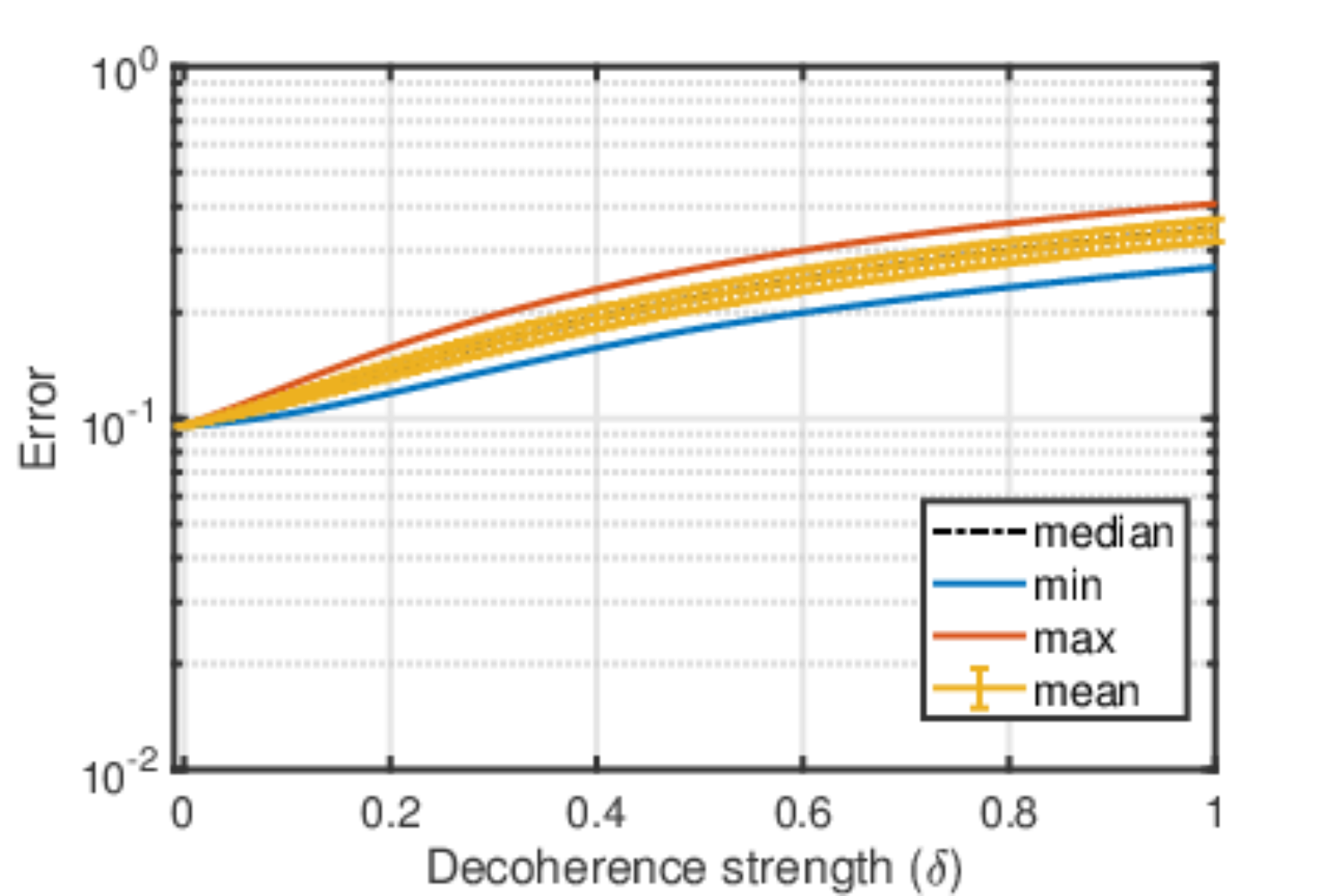}}
\caption{Minimum, maximum, median and mean error as function of a decoherence strength for two controllers for a five-spin ring optimized for transfer $1\to 2$.}\label{fig:N5_p12_best}
\end{figure}

To systematically study the sensitivity of different controllers to decoherence in the form of dephasing in the Hamiltonian basis, we sample the space of pure dephasing processes by generating a large set of lower triangular matrices $\gamma_{mn}^{(s)}$ of size $N$, the system dimension, with entries in $[0,1]$, randomly drawn from a uniform distribution.  A set of $1000$ dephasing operators was then generated by eliminating trial dephasing matrices that violate physical constraints~\cite{PhysRevA.86.012121}.  Each dephasing matrix $(\gamma_{k\ell}^{(s)})$ is then normalized
\begin{equation}
   \bar{\gamma}_{k\ell}^{(s)} = \gamma_{k\ell}^{(s)} / \sum_{1<k\le N, 1\le \ell<k} |\gamma_{k\ell}^{(s)}|
\end{equation}
and a decoherence strength parameter $\delta \in [0, 1]$ introduced.

For a given initial state $\rho_0$ and controller $D$, the output state $\rho^{(D,\delta,s)}(T)$ subject to dephasing is then calculated according to Eq.~\eqref{e:varrho_of_t_solution} with $\gamma_{k\ell}^{(\delta,s)} = -\delta \bar{\gamma}_{k\ell}^{(s)}$, and the distance from the output state $\rho^{(D)}(T)$ for the controller without dephasing is calculated as
\begin{equation}
  \epsilon(D,\delta,s) = \norm{\rho^{(D)}(T)-\rho^{(D,\delta,s)}(T)}.
\end{equation}
It is then straightforward to calculate the mean, standard deviation, minimum, maximimum and median of $\epsilon$ over all decoherence processes $s$ as a function of the decoherence strength parameter $\delta$ for each controller $D$.

\begin{figure}\center
\includegraphics[width=0.8\columnwidth]{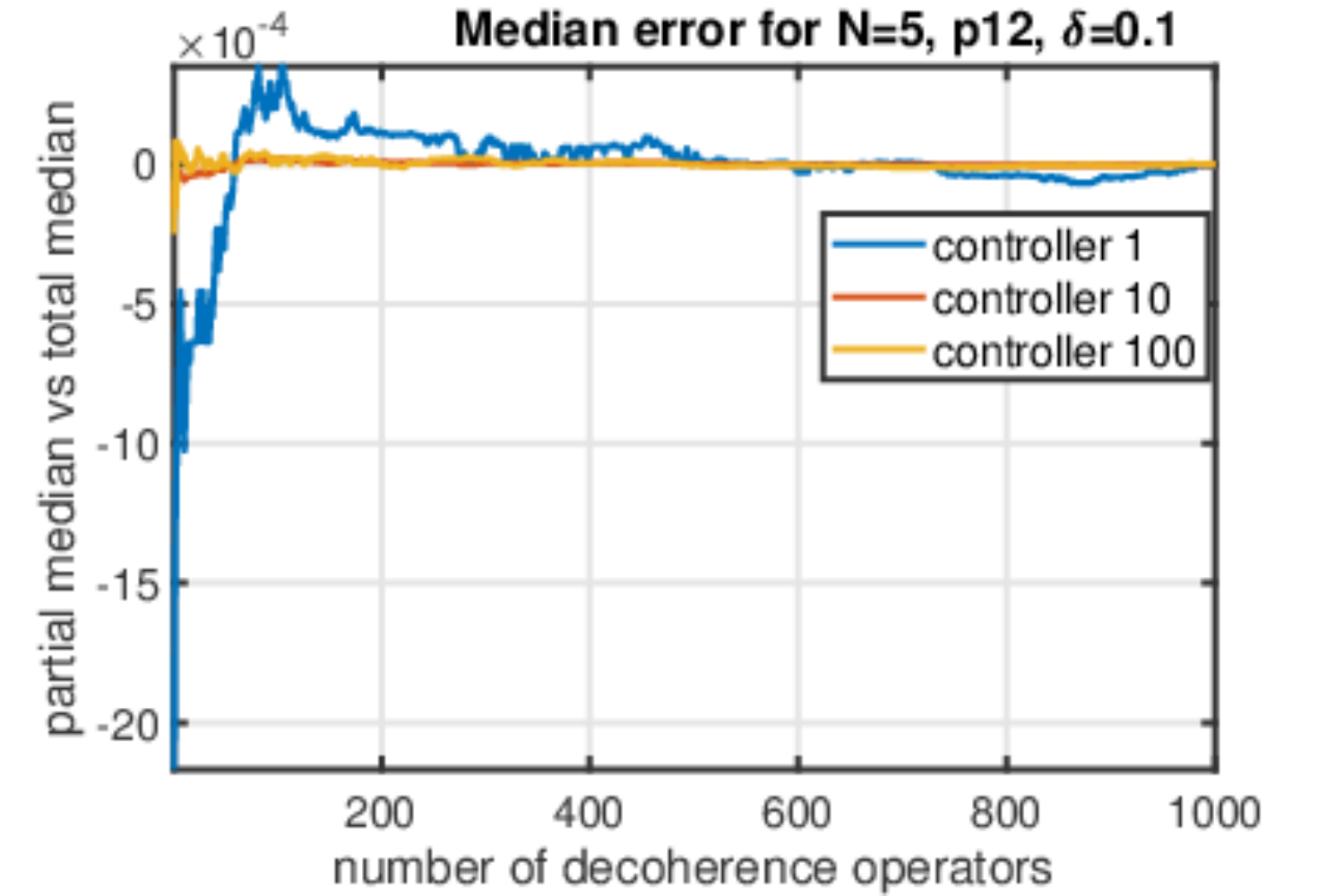}
\caption{Convergence of median error as a function of decoherence strength for three controllers 
from (1) high to (100) low fidelity for a five-spin ring optimized for transfer $1\to 2$.}\label{fig:convergence}
\end{figure}

Results for two controllers in Fig.~\ref{fig:N5_p12_best} show that the error increases faster for the good controller, as one might expect classically.  The deviation of partial median from the full median over $1000$ dephasing processes shown in Fig.~\ref{fig:convergence} shows that convergence is slower for higher fidelity controllers (controller 1) than for lower fidelity controllers but $1000$ dephasing processes appear to be sufficient for our test systems to estimate the error with a precision of about $5 \times 10^{-4}$.

\begin{figure} \center
  \vspace{-0.1in}
  \subfloat[Transfer $1\to 2$, 5-spin ring]{\includegraphics[width=0.75\columnwidth]{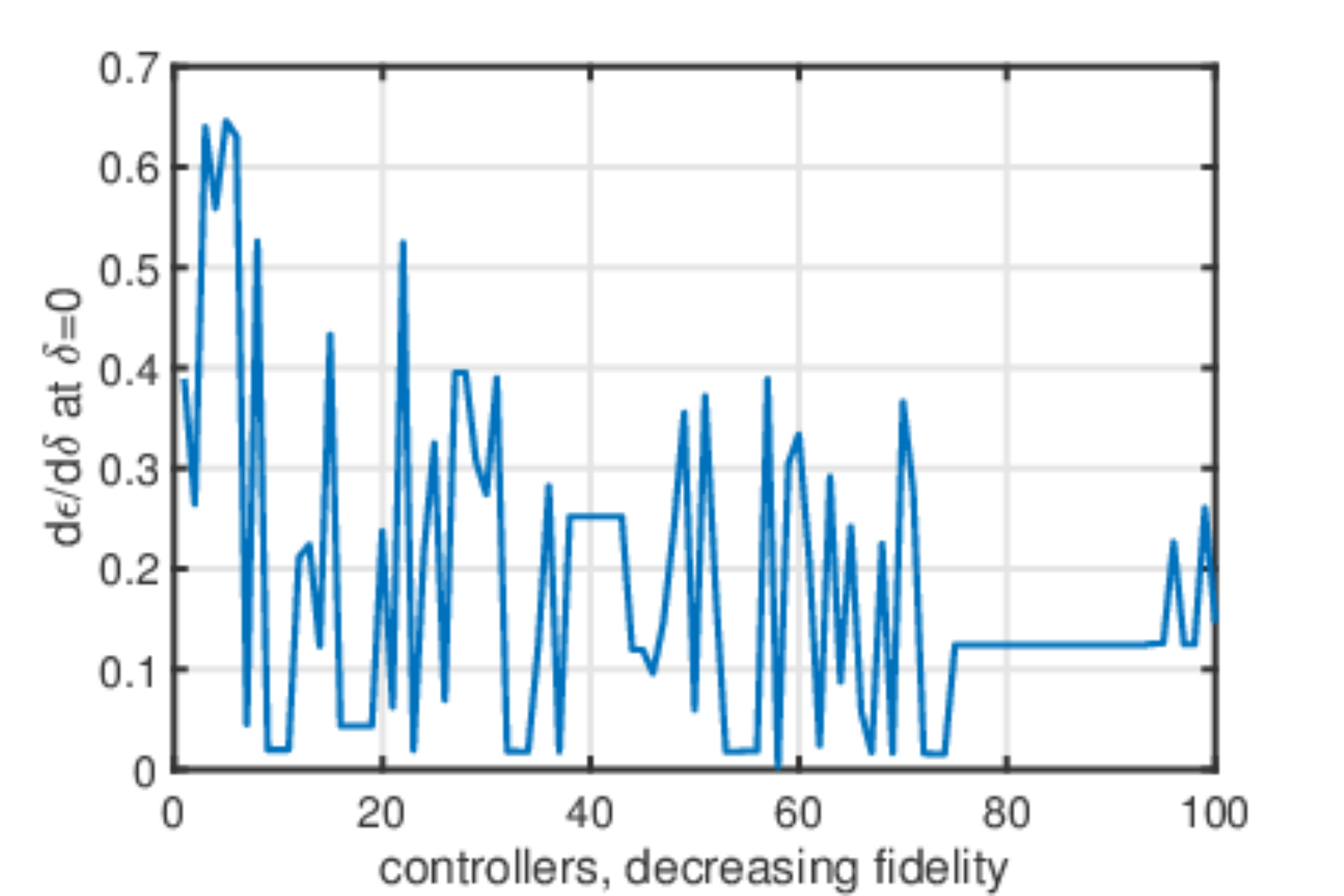}} \\[0pt]
  \subfloat[Transfer $1\to 3$, 5-spin ring]{\includegraphics[width=0.75\columnwidth]{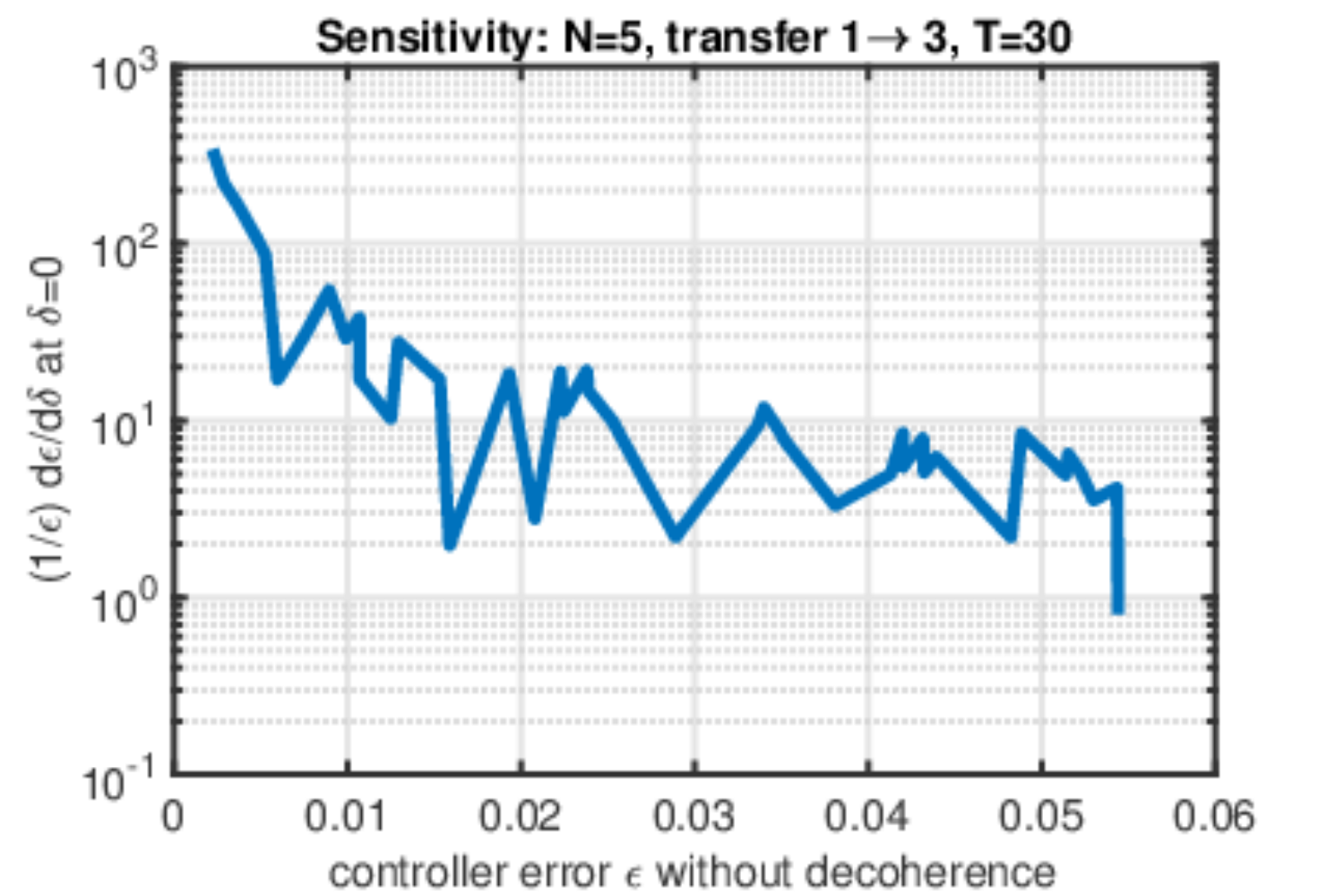}}
  \caption{Sensitivity (a) and log-sensitivity (b) for $100$ controllers.}\label{fig:N5_sensitivity1}
\end{figure}

\begin{figure} \center
  \subfloat[Transfer $1\to 2$, 5-spin ring]{\includegraphics[width=0.75\columnwidth]{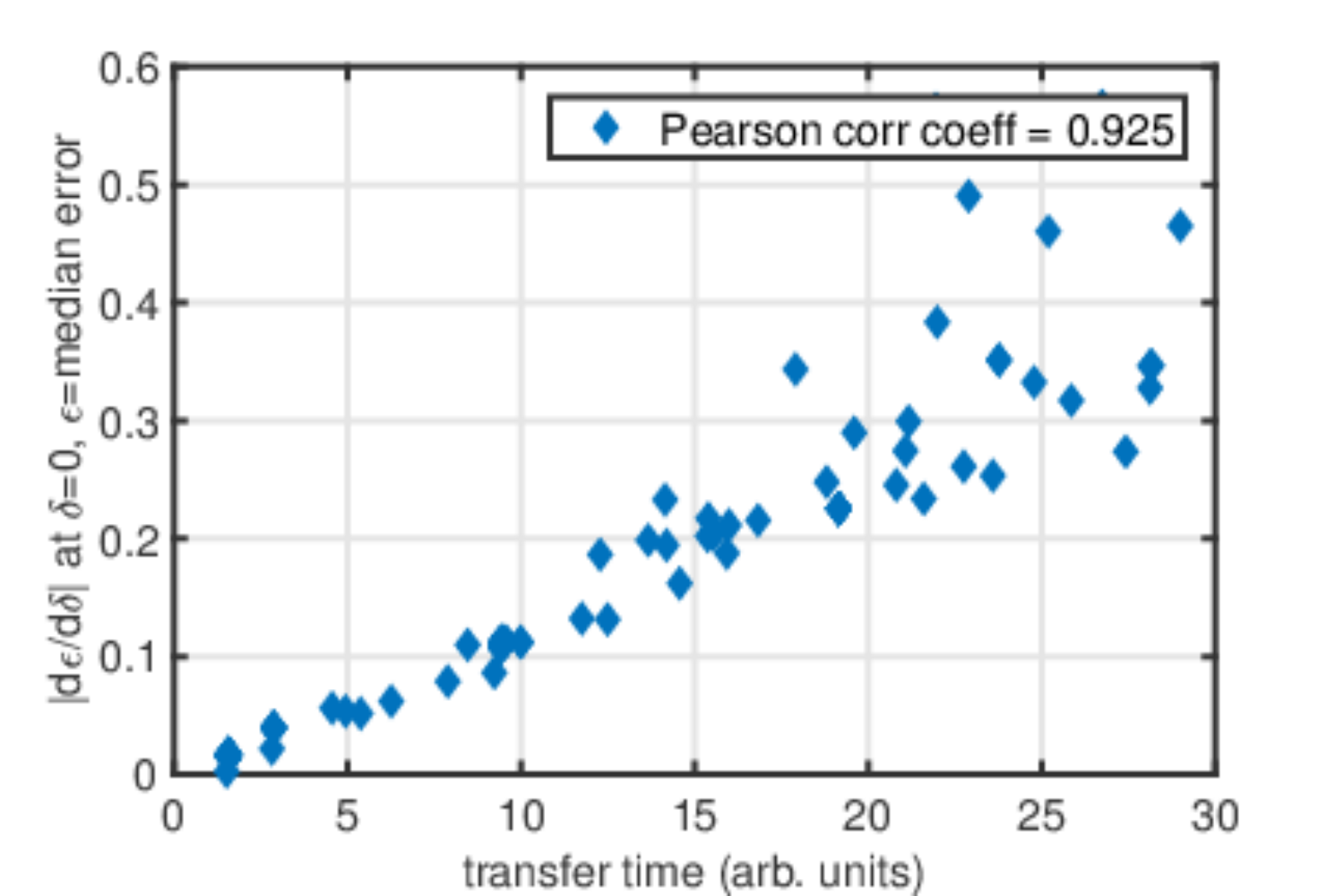}} \\
  \subfloat[Transfer $1\to 3$, 5-spin ring]{\includegraphics[width=0.75\columnwidth]{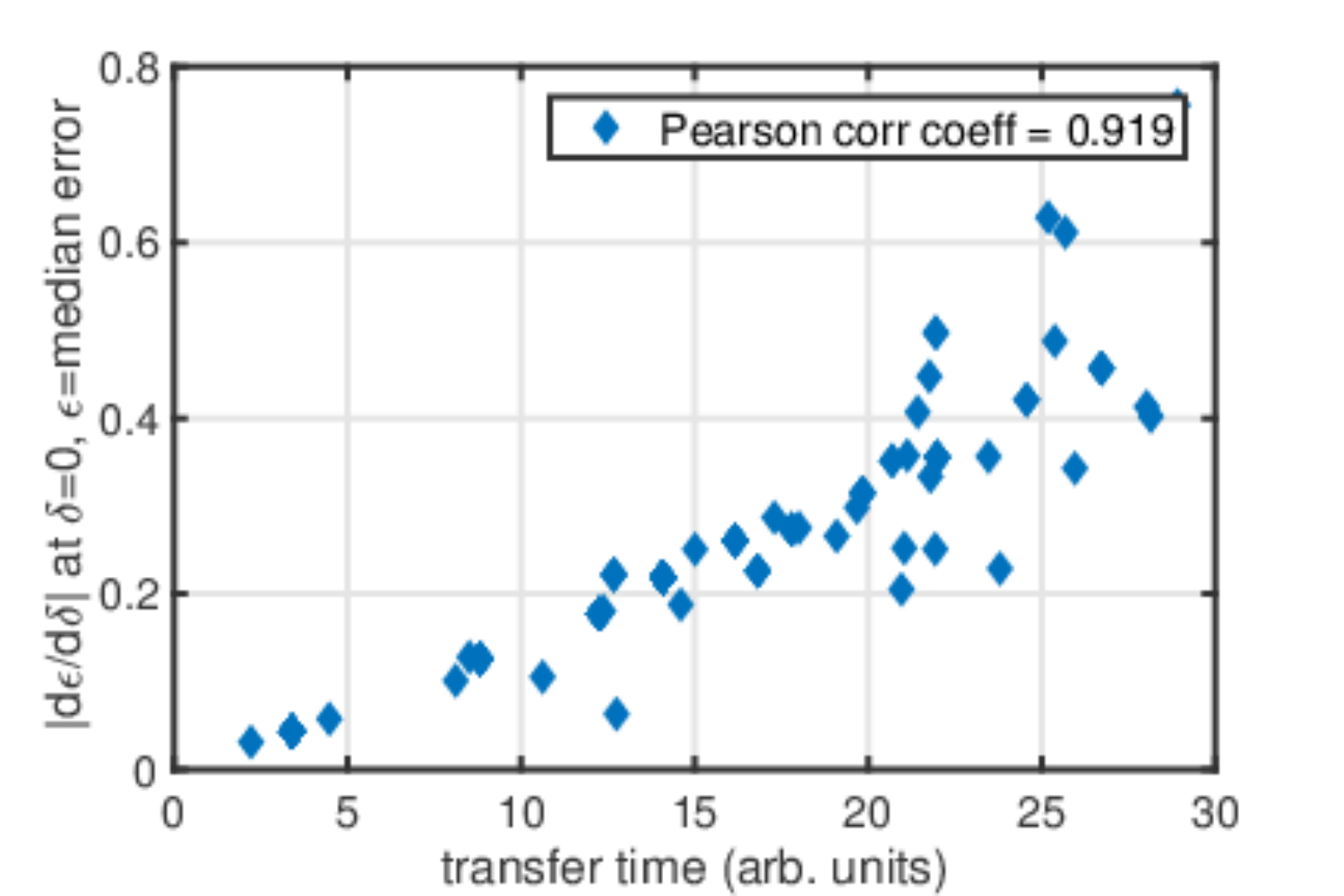}}
  \caption{Sensitivity $\eta(D)$ vs transfer time for $100$ controllers.}\label{fig:N5_sensitivity_vs_T}
\end{figure}

To quantify the sensitivity of the controller with regard to dephasing for small $\delta$, we use a finite-difference approximation of the derivative of $\eta(D) = d\epsilon(D,\delta)/d\delta$ at $\delta=0$, where $\epsilon(D,\delta)$ is the median error (over $1000$ dephasing processes).  Fig.~\ref{fig:N5_sensitivity1} shows the sensitivity of the controllers for $100$ controllers arranged in order of decreasing transfer fidelity, i.e., increasing eror.  There is a general tendency for the sensitivity to decrease, i.e., low-fidelity controllers are less sensitive to decoherence, as we might expect classically.  However, there are significant fluctuations, and some good controllers appear to be considerably less sensitive to decoherence than others. One might speculate that the sensitivity of the controllers to dephasing is related to the time required for the transfer, and indeed Fig.~\ref{fig:N5_sensitivity_vs_T}, showing the sensitivity $\eta(D)$ versus the transfer time $T$ for $100$ controllers indicates a strong linear correlation.

\begin{figure}\center
  \subfloat[Median transfer fidelities vs decoherence strength $\delta$]{\includegraphics[width=0.8\columnwidth]{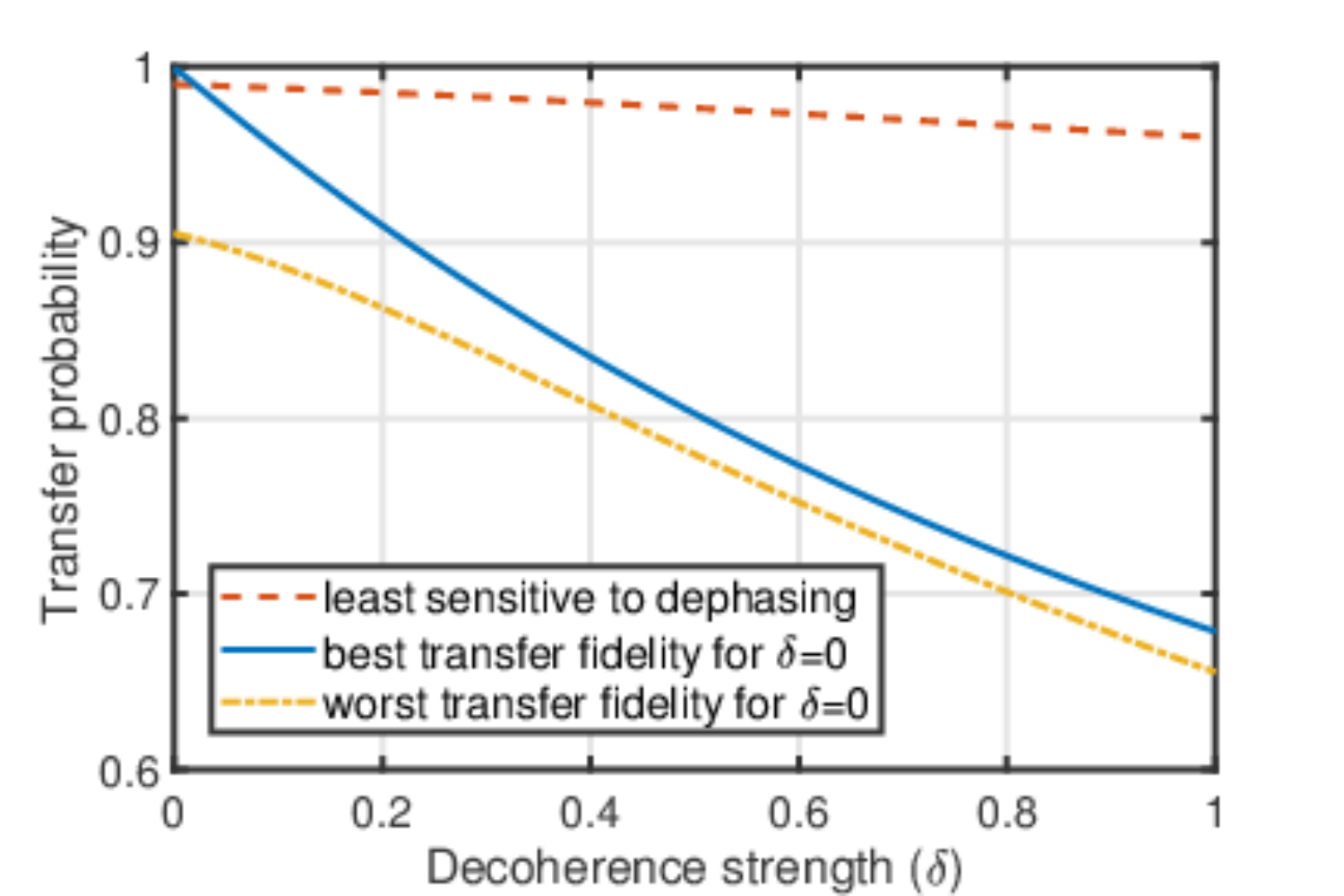}} \\
  \subfloat[Error distribution at $\delta=1$]{\includegraphics[width=0.8\columnwidth]{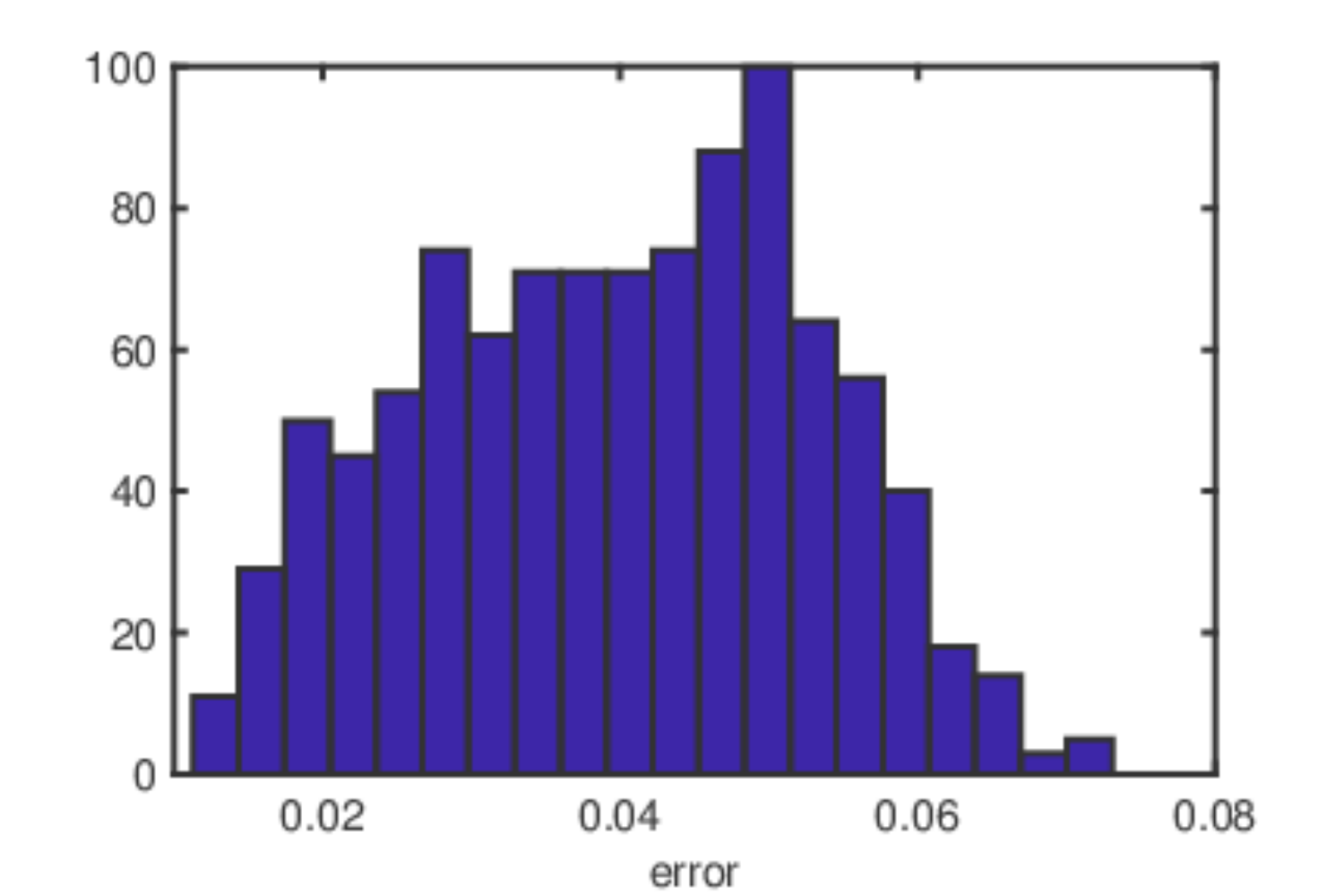}}
  \caption{Median transfer fidelity  for three different controllers (optimized for $1\to 2$ transfer in 5-spin ring) and error distribution for controller least sensitive to dephasing.}\label{fig:fidelity_compare}
\end{figure}

Finally, the median transfer fidelity for a five-spin ring as a function of the decoherence strength $\delta$ for three different controllers shown in Fig.~\ref{fig:fidelity_compare} suggests that there is generally significant potential to optimize the transfer fidelity in the presence of dephasing.  The median transfer fidelity for both the best and worst controller in terms of transfer fidelity at $\delta=0$ drops quite significantly, to 65\%-70\% for $\delta=1$, while for the most robust controller in the set, the initial $\delta=0$ transfer fidelity is slightly lower but the fidelity drops considerably less, still averaging around 95\% for $\delta=1$.  More importantly, the high median fidelity and the relatively narrow almost normal distribution of the error at $\delta=1$ for the controller least sensitive to dephasing, suggests that controllers could be optimized to be robust to generic dephasing when the dephasing rates are unknown.

\section{Conclusions}

We have modelled the effect of weak decoherence on coherent transport in spin networks using a Lindblad-type master equation and shown that under certain reasonable assumptions the spin dynamics can still be decomposed into subspace dynamics, which can be efficiently modelled.  We have shown that the steady state dynamics are closely related to long-term dynamic averages.  Numerical simulations suggest that controllers obtained for coherent transfer can still achieve good fidelities although higher fidelity coherent controllers tend to be more sensitive to decoherence---a ``classical" robustness feature.  Furthermore, there is considerable variation in terms of the sensitivity of controllers to decoherence, which suggests the potential to optimize the controllers to maximize the robustness of the transfer in the presence of weak decoherence.  Significantly, knowledge of the exact dephasing rates and processes does \emph{not} appear to be necessary to achieve robust transport.

Further work is necessary to derive optimal energy landscape controls for robust transfer of excitations in spin networks in the presence of decoherence, as well as uncertainties in the Hamiltonian and initial state preparation, which, due to the dependence of the decoherence on the Hamiltonian~\cite{domenico_CDC,singular_vs_weak_coupling}, also affect the decoherence processes and steady states, therefore magnifying their effect.


\begin{thebibliography}{10}
\providecommand{\url}[1]{#1}
\csname url@samestyle\endcsname
\providecommand{\newblock}{\relax}
\providecommand{\bibinfo}[2]{#2}
\providecommand{\BIBentrySTDinterwordspacing}{\spaceskip=0pt\relax}
\providecommand{\BIBentryALTinterwordstretchfactor}{4}
\providecommand{\BIBentryALTinterwordspacing}{\spaceskip=\fontdimen2\font plus
\BIBentryALTinterwordstretchfactor\fontdimen3\font minus\fontdimen4\font\relax}
\providecommand{\BIBforeignlanguage}[2]{{%
\expandafter\ifx\csname l@#1\endcsname\relax
\else
\language=\csname l@#1\endcsname
\fi
#2}}
\providecommand{\BIBdecl}{\relax}
\BIBdecl

\bibitem{QuantumSpintronicsReview}
D.D.~Awschalom, L.C.~Bassett, A.S.~Dzurak, E.L.~Hu, and J.R.~Petta,
  ``Quantum spintronics: Engineering and manipulating atom-like spins in
  semiconductors,'' \emph{Science}, vol. 339, no. 6124, pp. 1174--1179, 2013.

\bibitem{Glaser2015}
  S.J.~Glaser \emph{et~al.}, ``Training Schr{\"o}dinger’s cat: Quantum optimal control,''
  \emph{The European Physical Journal D}, vol.~69, no.~12, p. 279, 2015.

\bibitem{Edmond_IEEE_AC}
  S.~Schirmer, E.~Jonckheere, and F.~Langbein, ``Design of feedback control laws for information transfer in spintronics networks,''
  \emph{IEEE Trans.\ Autom.\ Control}, vol.~63, 2523-2536, 2018.

\bibitem{time_optimal}
  F.~Langbein, S.~Schirmer, and E.~Jonckheere, ``Time optimal information transfer in spintronics networks,''
  in \emph{Proc. 54th Conf.\ Decision Control}, Osaka, Japan, Dec 2015, pp. 6454--6459.

\bibitem{domenico_CDC}
  D.~D'Alessandro, E.~Jonckheere, and R.~Romano, ``Control of open quantum systems in a bosonic bath,''
  in \emph{Proc.\ 54th Conf.\ Decision Control}, Osaka, Japan, Dec 2015, pp. 6460--6465.

\bibitem{singular_vs_weak_coupling}
  D.~P. D'Alessandro, E.~Jonckheere, and R.~Romano, ``On the control of open quantum systems in the weak coupling limit,''
  in \emph{21st Int.\ Symp.\ Mathematical Theory Networks \&Systems ({MTNS})},
  Groningen, Netherlands, July 2014, pp. 1677--1684.

\bibitem{collective_dephasing}
Z.H.~Wang, Y.J.~Ji, Y.~Li, and D.L.~Zhou, ``Dissipation and decoherence induced by collective dephasing in a coupled-qubit system with a common bath,''
\emph{Phys.\ Rev.\ A}, vol.~91, 013838, 2015. 

\bibitem{Wang2010}
  X.~Wang and S.G.~Schirmer, ``Analysis of Lyapunov method for control of quantum states,''
  \emph{IEEE Trans.\ Autom.\ Control}, vol.~55, pp. 2259--2270, 2010.

\bibitem{Schirmer2010}
  S.~Schirmer and X.~Wang, ``Stabilizing open quantum systems by markovian reservoir engineering,''
  \emph{Phys.\ Rev.\ A}, vol.~81, 062306, 2010.

\bibitem{Motzoi2016}
F.~Motzoi, E.~Halperin, X.~Wang, K.~B. Whaley, and S.~Schirmer,  ``Backaction-driven, robust, steady-state long-distance qubit entanglement over lossy channels,'' \emph{Phys.\ Rev.\ A}, vol.~94, 032313, 2016.

\bibitem{formal_extended_final_value_Laplace}
  E.~Gluskin and S.~Miller, ``On the recovery of the time average of contionuous and discrete time functions from their {L}aplace and z-transform,'' 2011,  ar{X}iv:1109.3356v4 [math-ph].

\bibitem{statistical_control}
  E.~Jonckheere, S.~Schirmer, and F.~Langbein, ``Jonckheere-{T}erpstra test for nonclassical error versus log-sensitivity relationship of quantum spin network controllers,'' \emph{Int.\ J.\ Robust Nonlinear Control}, vol.~28, pp. 2383--2403, 2018.

\bibitem{new_eig_derivatives}
  F.~Wei, ``Efficient method for eigenvector derivatives with repeated eigenvalues,'' in \emph{33rd Structures, Structural Dynamics, Materials
Conf.}, Dallas, TX, 1992.

\bibitem{van_der_Aa}
  N.~van~der Aa, ``Computation of eigenvalue and eigenvector derivatives for a general complex-valued eigensystem,'' \emph{Electronic Journal of Linear Algebra}, vol.~16, pp. 300--314, 2007.

\bibitem{ssv_mu}
  E.~Jonckheere, S.~Schirmer, and F.~Langbein, ``Structured singular value analysis for spintronics network information transfer control,'' \emph{IEEE Trans.\ Autom.\ Control}, vol.~62, no.~12, pp. 6568--6574, 2017.

\bibitem{PhysRevA.86.012121}
  D.K.L.~Oi and S.G.~Schirmer, ``Limits on the decay rate of quantum coherence and correlation,'' \emph{Phys.\ Rev.\ A}, vol.~86, p. 012121, 2012.

\bibitem{soneil_mu}
  S.~O'Neil, E.~Jonckheere, S.~Schirmer,  F.~Langbein, ``Sensitivity and robustness of quantum rings to parameter uncertainty,'' in \emph{56th Conf. Decision Control}, Melbourne, Australia, Dec 2017, pp. 6137--6143.
\end{thebibliography}
\end{document}